# Dynamics of bound soliton states in regularized dispersive equations

**M.M. Bogdan, O.V. Charkina**

*B. Verkin Institute for Low Temperature Physics and Engineering of the National Academy of Sciences of Ukraine, 47 Nauky Ave., Kharkiv, 61103, Ukraine*

E-mail: m_m_bogdan@ukr.net , charkina@ilt.kharkov.ua

The nonstationary dynamics of topological solitons (dislocations, domain walls, fluxons) and their bound states in one-dimensional systems with high dispersion are investigated. Dynamical features of a moving kink emitting radiation and breathers are studied analytically. Conditions of the breather excitation and its dynamical properties are specified. Processes of soliton complex formation are studied analytically and numerically in relation to the strength of the dispersion, soliton velocity, and distance between solitons. It is shown that moving bound soliton complexes with internal structure can be stabilized by an external force in a dissipative medium then their velocities depend in a step-like manner on a driving strength.

## 1. Introduction

The soliton concept in applied science was formulated 35 years ago in the prominent review by Scott, Chu and McLaughlin [1]. Since the soliton research developed into both well-established mathematical and physical theories. They cover a wide range of problems beginning from complete integrability of nonlinear equations [2,3] up to applications of the soliton concept for explanation of nonlinear phenomena in various fields of condensed matter physics [4-7]. Topological defects and inhomogeneties such as dislocations in crystals, domain walls and vortices in magnets, quanta of magnetic flux (fluxons) in long Josephson junctions are a few examples of traditional physical objects which are described in terms of solitons in solid state physics.

Two pioneer works by Kosevich and Kovalev [8,9], devoted to nonlinear dynamics of one-dimensional crystals, initiated two novel directions in soliton investigations. In Ref. 8 the physical concept of a self-localized excitation was introduced for the first time. In the long-wavelength limit such an oscillating solitary wave corresponds to a breather, which is interpreted as the soliton-antisoliton bound state. The authors proposed a regular asymptotic procedure for constructing the self-localized oscillation [8,10]. In the short-wavelenth limit Kosevich and Kovalev predicted the existence of a self-localized oscillation with a frequency above the upper edge of a linear excitation spectrum. Later a high localization limit of these high-frequency soliton states were studied, and they were called the intrinsic localized modes or discrete breathers, which became a new concept in nonlinear lattice theory [11, 12].

In Ref. [9], which concerned crowdion dynamics in an one-dimensional anharmonic crystal, Kosevich and Kovalev established, to our knowledge for the first time, the existence of supersonic and radiationless motion of topological solitons in a highly dispersive nonlinear medium. The equations deduced in Ref. [9] generalize the Boussinesq equation for the case of the SG (SG) and $\varphi^4$-models:

$$u_{tt} - u_{xx} + (\alpha - \gamma u_x)u_x u_{xx} - \beta u_{xxxx} + F(u) = 0, \qquad (1)$$

where the external force is equal to either $F(u) = \sin u$ or $F(u) = u^3 - u$, respectively. The equation (1) with the sine force and $\alpha = 0$

$$u_{tt} - u_{xx} - \gamma u_x^2 u_{xx} - \beta u_{xxxx} + \sin u = 0 \tag{2}$$

is known nowadays as the Kosevich-Kovalev equations [13,14]. For the special choice of parameter $\gamma = 3\beta/2$, Kosevich and Kovalev found an exact solution describing the $2\pi$-kink moving with an arbitrary velocity [9]. One year later an integrable version of the equation was proposed by Konno, Kameyama, and Sanuki [15]:

$$u_{xt} + \frac{3}{2}\beta u_x^2 u_{xx} + \beta u_{xxxx} - \sin u = 0 \tag{3}$$

and this fact could explain formally the existence of the exact kink solution in Eq. (2). However, the significance of the fact of radiationless motion of topological solitons in highly dispersive media was realized after many years. In 1984 Peyrard and Kruskal showed numerically the existence of a stable moving $4\pi$-soliton in the highly discrete SG model [16]:

$$\frac{\partial^2 u_n}{\partial \tau^2} + 2u_n - u_{n-1} - u_{n+1} + \frac{1}{d^2}\sin u_n = 0, \tag{4}$$

where $d$ is the discreteness parameter. They tried to explain the formation of the bound state of two identical kinks by exploiting the fact of the presence of the Peierls potential in the lattice model. However, in work [17] it has been found that the radiationless motion of such a soliton complex can be described explicitly by the exact $4\pi$-soliton solution in the framework of the dispersive SG equation with a fourth spatial derivative, i.e. Eq.(2) with $\gamma = 0$:

$$u_{tt} - u_{xx} - \beta u_{xxxx} + \sin u = 0 \tag{5}$$

which is obtained as the long-wavelength limit of the Eq. (4). Almost simultaneously the topological bound soliton states were found numerically in a continuous nonlocal SG model describing long Josephson junctions [18]. These facts of existence of the multikink bound states in discrete and continuous systems were generalized as a universal phenomenon and led to the concept of the soliton complexes formed by strongly interacting kinks in highly dispersive media [19-21]. Physically such two-kinks states correspond, e.g., to a moving defect consisting of two neighboring dislocation half-planes, or to a narrow $360^0$ magnetic domain wall, which arises even in the absence of magnetic field, or to a bound pair of fluxons in a long Joshephson junction.

There are several approaches to explaining the mechanisms of formation of the bound soliton states. The internal structure of the soliton complexes can be studied in detail in models that lead to piecewise-linear equations with strong dispersion [20-22]. In this case the stationary states can be constructed as a superposition of two quasi-solitons possessing spatial periodic tails as asymptotics which cancel each other exactly for the composed complex by imposing some interference condition. Since these bound solitons occur in resonance with the linear spectrum waves they were called embedded solitons [23,24]. The effect of the dispersion can be extracted already from the dispersion relations of the corresponding linearized equations [25,26]. However, a key circumstance for a complex to arise appears to be the influence of strong dispersion as a factor leading to complication of the internal structure of solitons, starting from a kink level [27]. Taking the interaction of such flexible kinks into account allows one to describe quantitatively the conditions of formation of a soliton complex [19,21,28].

The picture of soliton complex formation becomes much more diverse when one considers internal dynamics of kinks, nonstationary motion of complexes, and the conditions of their formation and stability, depending on different physical factors including the influence of dissipative and external forces [29-31]. The present paper is devoted to investigation of this circle of tasks, concentrating on a single kink propagation and especially on the bound soliton states of both types, soliton complexes and breathers, covering essentially nonlinear dynamics of the strongly dispersive SG model.

The paper is organized as follows. Section 2 introduces regularized dispersive equations and some their dynamical properties. Section 3 addresses the nonstationary dynamics of a single $2\pi$-kink in the entire range of the dispersive parameter. Section 4 is devoted to analysis of complex formation and its stability conditions. Section 5 deals with the breather dynamics. Section 6 addresses the influence of dissipation and external forces on stabilization of the soliton complexes with internal structure. Finally, the results obtained are summarized in Sec. VII.

## 2. The regularized dispersive SG equations

To investigate analytically and numerically the nonstationary dynamics of kinks and their bound states we use the regularized dispersive SG equation with a fourth-order spatio-temporal derivative [19-21]:

$$u_{tt} - u_{xx} - \beta u_{xxtt} + \sin u = 0 \qquad (6)$$

where $\beta$ is a dispersive parameter. This equation has an advantage in comparison with Eq. (5) because it does not contain an artificial instability of states $u = 0, 2\pi, 4\pi...$ with respect to a short-wavelength excitation. The idea of the regularization of dispersive equations belongs to Boussinesq, who first proposed to use a mixed spatio-temporal derivative instead of the fourth spatial derivative for the shallow-water waves equations [4]. Such a replacement was justified in the lattice theory by Rosenau [32] for models with nonlinear interactions between atoms. Boussinesq's idea was applied to the SG and double SG equations with higher dispersion in Refs. [19-21]. At present this approach is actively being used for analytical description of discreteness effects [33-35]. With respect to the original discrete models the accuracy of this replacement can be easy to estimate. In the long-wavelength limit ($d \gg 1$) after introducing a coordinate $x = n/d$ the second difference is replaced as $u_{n-1} + u_{n+1} - 2u_n \approx u_{xx} + \beta u_{xxxx}$, where $\beta = 1/12d^2$. If one expresses the second derivative from Eq.(5) as $u_{xx} = u_{tt} - \beta u_{xxxx} + \sin u$ and inserts it in the fourth derivative and keeps terms which are linear with respect to $\beta$, one obtains the equation:

$$u_{tt} - u_{xx} + \sin u - \beta u_{xxtt} - \beta (\sin u)_{xx} = 0 \qquad (7)$$

Hence, to approximate Eq.(5) by Eq.(6), one needs to take into account the term with $-\beta(\sin u)_{xx}$. It would be expected that the form of static kinks would be different for all the Eqs. (4)–(7), depending on a value of $\beta$. Curiously, it appears that Eq.(5) does not possess a static $2\pi$-kink solution satisfying the boundary conditions $u(-\infty) = 0$ and $u(\infty) = 2\pi$ at all [21]. At the same time, exact static kink and moving complex solutions exist simultaneously in Eq. (6). Therefore, many problems of kink and complex dynamics can be solved analytically in the framework of this equation. In particular, the spectral problem for linear excitations of the static kink has been solved completely [14,29]. Thus Eq. (6) has an exact static kink solution for arbitrary $\beta$, which coincides with a kink of the usual SG equation:

$$u_{2\pi}(x) = 4\arctan\exp(x) \tag{8}$$

The kink solution of Eq. (7) can be found in implicit form using the first integral:

$$\frac{du}{dx} = \frac{2\sin(u/2)}{1+\beta\cos u}\sqrt{1+\beta\cos^2(u/2)} \tag{9}$$

It appears that even when the discreteness parameter $d=1$ and hence $\beta=1/12$, the static kink solutions for the discrete equation (4) and the continuous Eqs. (6) and (7) differ very slightly (see Fig.1). This justifies the use Eq. (6) instead Eqs. (5) and (7) to explain qualitatively a majority of effects which are inherent in the discrete model (4) but in reality arise due to the higher-order dispersion.

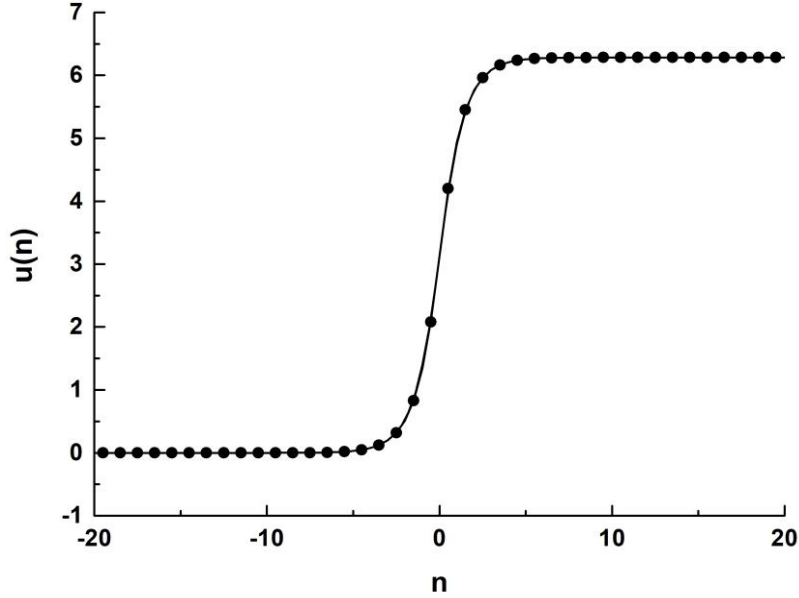

**Fig.1.** *Comparing static kink profiles for a discrete Eq. (4) and Eqs. (6) and (7) for $d=1$ ($\beta=1/12$). Continuous solutions are undistiguishable.*

The equation (6) can be derived from the Lagrangian:

$$L = \int \frac{1}{2}\left[u_t^2 - u_x^2 + \beta u_{xt}^2 - (1-\cos u)\right]dx. \tag{10}$$

Using the expression (10) it is easy to find the first integrals, total energy and momentum:

$$E = \int \frac{1}{2}\left[u_t^2 + \beta u_{xt}^2 + u_x^2 - (1-\cos u)\right]dx, \tag{11}$$

$$P = \int_{-\infty}^{\infty} u_x(u_t - \beta u_{xxt})dx. \tag{12}$$

Note that first two terms in the Eq. (11) give the kinetic energy, and therefore the higher-order dispersion in the regularized equations contributes to the kinetic energy, whereas in the case of Eq. (5) it produces an additional contribution to the potential energy [21].

The spectrum of linear excitations for Eq. (6) can be found exactly both for the cases of a homogenous ground state and in the presence of the static kink (8) [14,29]. The dispersion relation for continuous waves takes the form:

$$\omega(k) = \sqrt{(1+k^2)/(1+\beta k^2)} \tag{13}$$

This spectrum has the peculiarity of being bounded in frequency not only from below but also from above. This property makes it similar to the spectrum of the initial discrete model (4). Moreover, it simply coincides with the spectrum of the SG model with a nonlocal interaction [18]. In the case of a kink there exists a discrete spectrum of internal modes of oscillations [29], the number of which becomes infinite when $\beta \to 1$ while the continuous spectrum degenerates to one frequency $\omega_0 = 1$.

Finally, it is remarkable that Eq.(5) and (6) have exact solutions describing a moving $4\pi$-soliton complexes. For Eq. (6) the moving bound state of strongly coupled kinks has a form:

$$u_{4\pi}(x,t) = 8\arctan\left\{\exp\left(\frac{x-V_0 t}{l_0}\right)\right\}. \tag{14}$$

The velocity $V_0$ of such a complex, its effective width $l_0$ and its energy $E_0$ are specified functions of the parameter $\beta$:

$$V_0(\beta) = \sqrt{1+\frac{\beta}{3}} - \sqrt{\frac{\beta}{3}}, \qquad l_0 = \left(3\beta V_0^2\right)^{1/4}, \qquad E_0 = 32\left(l_0^{-1} - \frac{l_0}{9}\right). \tag{15}$$

In the next two sections we discuss dynamical properties of a single kink and specify conditions of soliton complex formation.

### 3. Dynamics of a kink in the dispersive SG model

Internal oscillations of the static kink of the regularized Eq. (6) have been studied theoretically, and main features of its nonstationary motion have been revealed numerically [14,29-31]. Here we present analytical approaches to the kink dynamics. One of them consists in the application of a perturbation theory for the case of a weak dispersion (small $\beta$). In this limit (the) dynamical properties of a kink would be expected to be similar to those in the usual SG equation. The latter has a moving kink $u_{2\pi}(z)$ obtained from the expression (8) by the Lorentz transformation of coordinates: $z = (x-Vt)/\sqrt{1-V^2}$. Therefore, one can seek a solution of Eq. (6) in the form:

$$u(x,t) = u_{2\pi}(z) + u_1(z,\tau) \tag{16}$$

where $\tau = (t-xt)/\sqrt{1-V^2}$, and the function $u_1$, a small addition to the kink form, obeys the linearized equation

$$\left(\frac{\partial^2}{\partial \tau^2} + L\right)u_1 \equiv u_{1\tau\tau} - u_{1zz} + \left(1 - \frac{2}{\cosh^2 z}\right)u_1 = \beta\left(u_{2\pi}(z) - u_1\right)_{xxtt} \tag{17}$$

In the first approximation one has to neglect the term $u_1$ in the right-hand side of Eq.(17). Then it is easy to find a partial solution of the equation:

$$\Delta u(z) = \alpha\left(3\frac{\sinh z}{\cosh^2 z} - \frac{z}{\cosh z}\right), \qquad \alpha = \frac{\beta V^2}{(1-V^2)^2} \qquad (18)$$

A general solution can be written as $u_1(z,\tau) = \Delta u(z) + v(z,\tau)$ where $v(z,t)$ is a solution of the homogeneous part of Eq.(17) (without the right-hand side). This solves the evolution problem of the SG kink in the dispersive system for the case of small $\alpha$. Indeed suppose that at the initial moment $u(z,0) = u_{2\pi}(z)$ and $u_{1\tau}(z,0) = 0$. This means that $v(z,0) = -\Delta u(z)$ and $v_\tau(z,0) = 0$. Owing to the knowledge of eigenfunctions of the operator $L$, one can solve completely the initial problem for the function $v(z,t)$:

$$v(z,\tau) = -\frac{\alpha}{4}\int_{-\infty}^{\infty}\frac{1+3k^2}{1+k^2}\cos\left(\sqrt{1+k^2}\,\tau\right)(\cos kz\,\text{th}\,z + k\sin kz)\frac{1}{\cosh\frac{\pi}{2}k}dk. \qquad (19)$$

This addition to the kink form describes decaying oscillations of the effective kink width which correspond to the SG quasimode. The time-dependent addition $\Delta\kappa(\tau)$ to the reverse effective kink width $\kappa_0 = 1+\alpha$ is easily found from Eq.(19):

$$\Delta\kappa(\tau) = \frac{1}{2}\frac{\partial v(z,\tau)}{\partial z}\bigg|_{z=0} = -\frac{\alpha}{8}\int_{-\infty}^{\infty}(1+3k^2)\frac{\cos\left(\sqrt{1+k^2}\,\tau\right)}{\cosh\frac{\pi}{2}k}dk, \qquad (20)$$

and its temporal behavior (Fig.2) repeats entirely the kink velocity modulation found numerically [29]. The power spectrum of the oscillation reveals Rice's frequency value $\omega_R = 2\sqrt{3}/\pi$ [36].

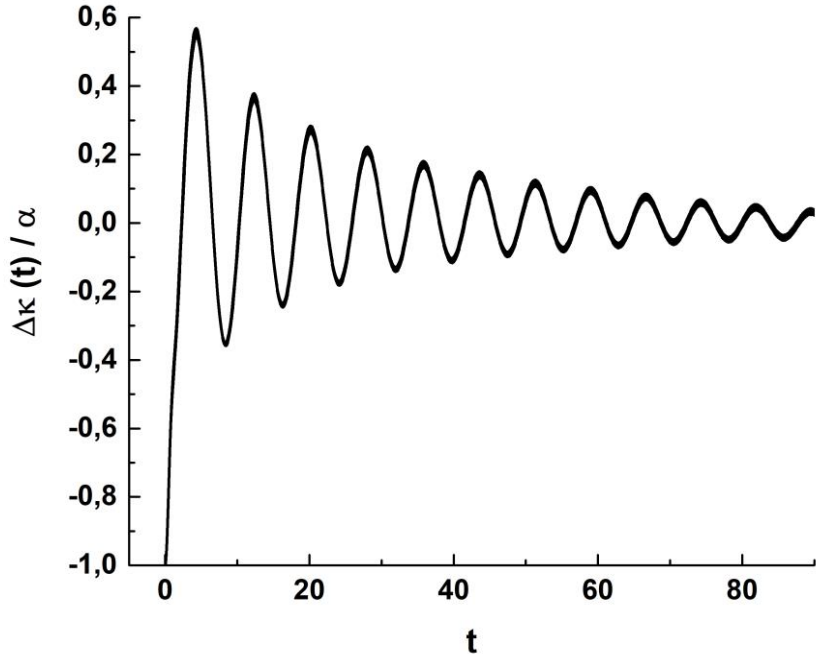

**Fig.2.** *Decaying oscillations of the addition to the reverse effective kink width during motion in the case of small $\beta$.*

Thus the perturbation theory predicts that the initial SG kink has to evolve into a steady moving profile $u_K(z) = u_{2\pi}(z) + \Delta u(z)$. However, it is known [21] that the equation for stationary waves

$$u_{zz} + \alpha u_{zzzz} - \sin u = 0, \qquad (21)$$

does not possess an exact solution for a moving $2\pi$-kink, although one can formally find the first terms in an asymptotic series for such a solution, which coincide with Eq.(18). The paradox is resolved by noting that the solution $u_K(z)$ can be expressed as superposition of two $\pi$-kinks in the form

$$u_K(z) = 2\left\{\arctan\exp\left[\left(1-\frac{\alpha}{2}\right)z + i\sqrt{3\alpha}\right] + \arctan\exp\left[\left(1-\frac{\alpha}{2}\right)z - i\sqrt{3\alpha}\right]\right\}. \qquad (22)$$

which prompts the ansatz for an adiabatic approach to the $2\pi$-kink dynamics. The nonstationary evolution of the kink at small $\alpha$ reduces to decaying collective oscillations of the effective kink width and velocity, with a consequent growth of a kink steepness and a slow energy loss due to the radiation emission. With increasing dispersive parameter a noticeable oscillating kink tail appears, and this phenomenon can be described by the following ansatz:

$$u_{Kr}(z,t) = u_K(z) + a \cdot [1 - \tanh(z)] \cdot \sin(k_0(z - vt)), \qquad (23)$$

where the second term corresponds to radiation on the wake of the kink. We have carried out a numerical modeling of the dynamics of kinks and soliton complexes (details of the numerical scheme can be found in Ref. 31). Results of the simulations for small $\beta$ are in a good relation with expression (22) and (23) and confirm entirely theoretical predictions. For large enough parameter $\beta$ and the initial velocity $V_{in}$ a moving kink emits the breather, as shown in Fig.3.

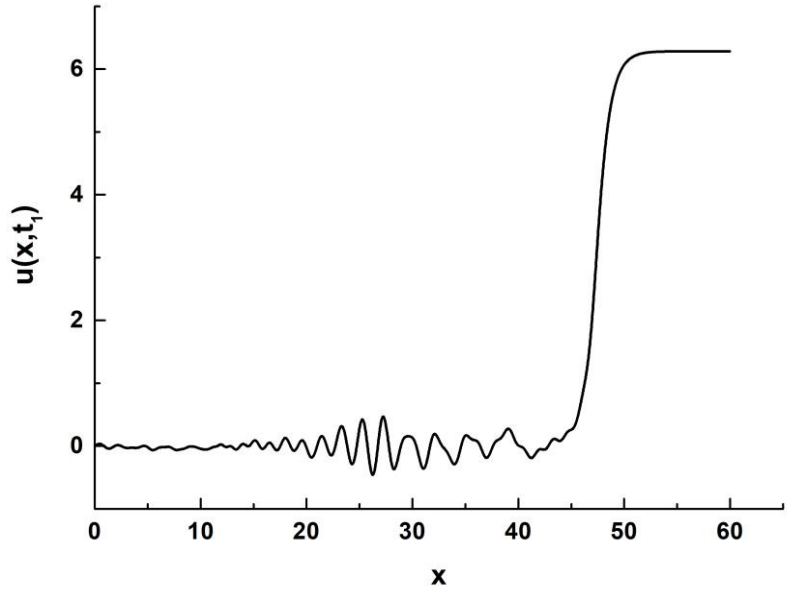

**Fig.3.** *Evolution of a fast kink generating a breather on its wake for $\beta = 1/6$ and $V_{in} = 0.86$ at $t_1 = 70$.*

## 4. Kinks interaction and formation of soliton complexes

An analytical approach to the description of the soliton-complex formation in dispersive equations was proposed in (Refs.) [21,29]. It is based on the use of the collective variable ansatz, which is constructed by taking into account the translational and internal degrees of freedom of a soliton as well as interactions between solitons and solitons with radiation. Now, using results of previous section, we can specify the form of ansatz:

$$u_{wb}(x,t) = u_K(\xi + R) + u_K(\xi - R) + f_b(\xi,t) \cdot (1 - \tanh(\xi)) \qquad (24)$$

Here first two terms are the superposition of kinks and the last term describes a small-amplitude breather $f_b(\xi,t) = a\sin(\Omega t - k(\xi - \xi_0))/\cosh(\varepsilon(\xi - \xi_0))$ or radiation emitting. It turns out that the condition of complex formation of closely sited solitons can be found from the energy expression of a pair of strongly interacting solitons without taking into account the breather or radiation [19,21,28]. Now we use this approximation for the description of the regularized SG system. Thus we suppose that the complex dynamics can be considered in the framework of the soliton ansatz

$$u_{kk}(x,t) = 4\arctan(\exp(\xi + R)) + 4\arctan(\exp(\xi - R))) \qquad (25)$$

which is prompted by the form of a generalization of the exact solution in Eq. (14). Here $\xi = \kappa(x - X(t))$ and $X(t)$, $\kappa(t)$, and $R(t)$ are functions of time. Functions $\kappa(t)$ and $X(t)$ describe the changing of the effective width of solitons and their translational motion, respectively. The function $R(t)$ corresponds to the changing separation between solitons, which is defined obviously as $L = 2R/\kappa$. Let the distance between solitons be small. Inserting the ansatz into Eqs. (11) and (12), we find the effective Lagrangian for two interacting solitons in the strongly dispersive medium:

$$L = 16\left\{\frac{\kappa_t^2}{\kappa^3}\left[\frac{\pi^2}{12} - R^2\left(\frac{\pi^2}{36} - \frac{2}{3}\right)\right] - \frac{\kappa_t}{\kappa^2}RR_t + \kappa\left(X_t^2 - 1\right)\left(1 - \frac{R^2}{3}\right) - \frac{1}{3\kappa}\left(1 + \frac{R^2}{5}\right) + \right.$$
$$\left. + \frac{\beta}{2}\left\{\frac{\kappa_t^2}{\kappa}\left[\left(\frac{\pi^2}{18} + \frac{2}{3}\right) - R^2\left(\frac{7\pi^2}{90} - \frac{2}{3}\right)\right] - 2\kappa_t RR_t + \frac{2}{3}\kappa^3 X_t^2\left(1 - \frac{7R^2}{5}\right)\right\}\right\} \qquad (26)$$

Analysis of Eq. (26) shows that the soliton complex is stable for high values of its velocity and a small distance between composite kinks due to the effective kinks attraction. For value of velocity much larger than the velocity of stationary motion the complex also dissociates in the manner shown in Fig.4.

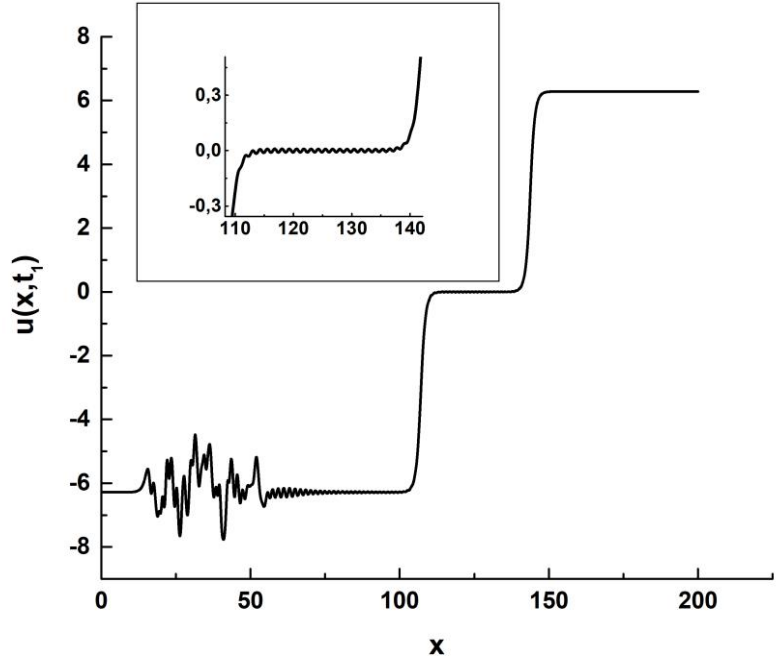

**Fig.4.** *Decay of a soliton complex for $\beta = 1$, $V_{in} = 0,9$ and $t_1 = 500$. The first kink moves with a almost constant velocity about $V_1 = 0.152$. Behind the second kink are breather modes. The inset shows the spatial modulation of the field between kinks on an expanded scale.*

### 5. Breather properties in the regularized dispersive SG equation

The form of static breather can be found analytically as an asymptotic series using the Kosevich-Kovalev scheme of construction of the self-oscillation solution [8]:

$$u(x,t) = A(x)\sin \omega t + B(x)\sin 3\omega t + C(x)\sin 5\omega t + \ldots . \qquad (27)$$

For the main harmonics with a frequency $\omega$ which is close to the linear spectrum lower edge, i.e. for $\varepsilon = \sqrt{1-\omega^2} \ll 1$, one obtains the following effective equation:

$$\psi_{tt} - \psi_{xx} + \psi - \beta \psi_{xxtt} - \frac{1}{8}|\psi|^2 \psi = 0. \qquad (28)$$

Here a complex function $\psi(x,t)$ determines the solution $u(x,t) = \text{Re}(\psi(x,t))$ in the first approximation with respect to the small parameter $\varepsilon$. Seeking the solution of Eq.(28) in the form $\psi = f(x)\exp(i\omega t)$ we derive the nonlinear ordinary equation:

$$(1-\beta\omega^2)f_{xx} - (1-\omega^2)f + \frac{1}{8}f^3 = 0, \qquad (29)$$

which gives the coordinate dependence of the harmonic amplitude as a usual soliton profile:

$$f = \frac{4\varepsilon}{\cosh \kappa x}, \quad \kappa^2 = \frac{1-\omega^2}{1-\beta\omega^2} . \qquad (30)$$

However, one can see a new feature of the breather, which consists in vanishing the effective width dependence on the amplitude $\varepsilon$ in the limit $\beta \to 1$. In fact, it appears that in this case the amplitude of the breather is no longer a constant but a slowly time-oscillating function. This results in the main frequency splitting and the complex breather behavior showing in Fig.5. Such behavior is similar to dynamical properties of breathers in discrete and nonlocal SG models [37,38]. Last we have found that a single breather motion is accompanied by a small breather bursting process and emitting radiation, as shown in Fig.6. As we have seen in previous sections, the excitation of breather modes plays a crucial role in the kink and soliton complex dynamics in the case of a strong dispersion.

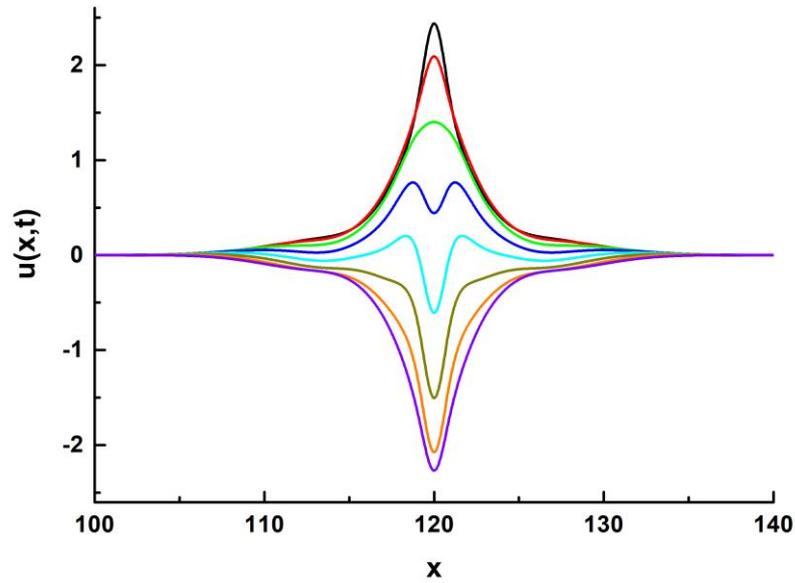

**Fig.5.** *A half-period evolution of a static breather at $\beta = 0.9$.*

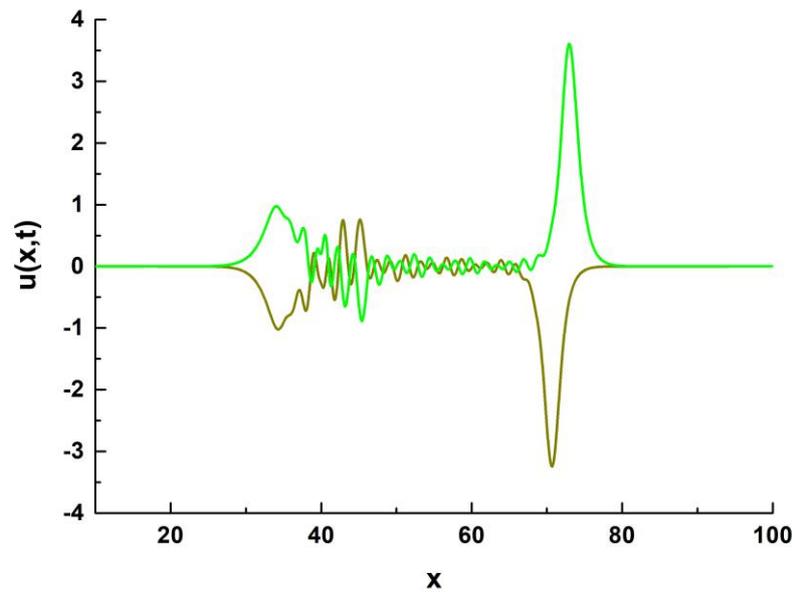

**Fig.6.** *Two moving breather profiles separated by a half-period in time at $\beta = 0.9$.*

## 6. Stabilization of soliton complexes by driving forces in dissipative media

Finally, we have investigated the influence of external forces and dissipation on the dynamics of soliton complexes. For this purpose we add a dissipation term $\lambda u_t$ and a driving force $f_0$ to the right-hand side of Eq. (6)

$$u_{tt} + \lambda u_t - u_{xx} - \beta u_{xxtt} + \sin u = f_0 \qquad (31)$$

The term $f_0$ in the right-hand side corresponds, for example, to the bias current in a long Josephson junction. The result of a numerical modeling are presented in Fig.7 for $4\pi$-complex profiles and in Fig.8 for their step-like velocity dependences on the driving force strength (one can compare this result with the velocity-force dependence for a single $2\pi$-kink in the discrete SG model [39]). The parameters are chosen as follows: $\lambda = 0.1$ and six sequential values of $f_0$ from −0.1 to −0.35. It turns out that the driving force under conditions of dissipation permits stabilization not only of the soliton complex but also of its "excited" states with internal structures. For waves of stationary profile, the derivatives $u_t$ and $u_x$ are proportional to each other, and both have the form of closely spaced double peaks. These derivatives are directly related to experimentally measurable quantities, in particular, the voltage $U \sim u_t$ and magnetic field $H \sim u_x$ in the case of a long Josephson junction, and in a crystal with dislocations the derivative $u_x$ determines the elastic deformation of the medium. In conclusion, we note that the possibility of observing multisoliton excitations in long Josephson junctions was demonstrated quite some time ago [40].

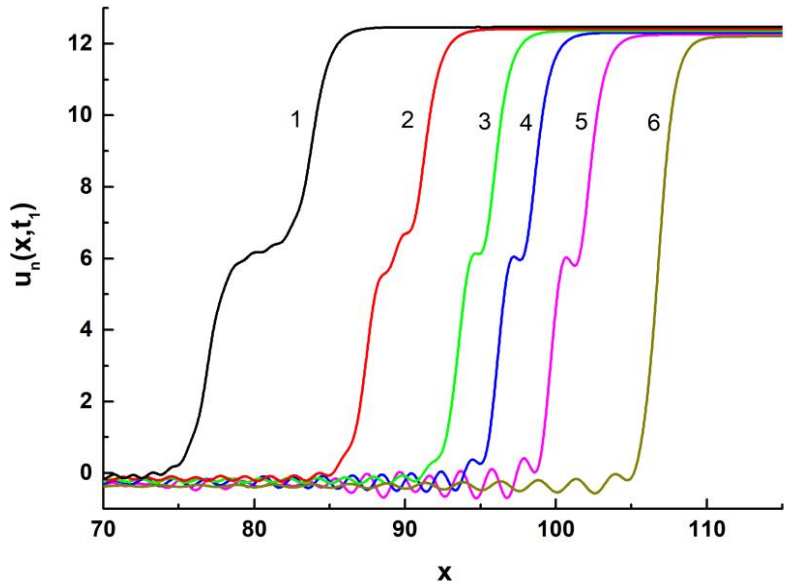

**Fig.7.** *Propagation of stable soliton complexes with an internal structure under the influence of external forces and in the presence of dissipation. The coefficient $\lambda = 0.1$ and the curves with numbers $n = 1,...,6$ corresponds to $f_0 = -0.05(1+n)$ and to the same instant of time $t_1 = 100$.*

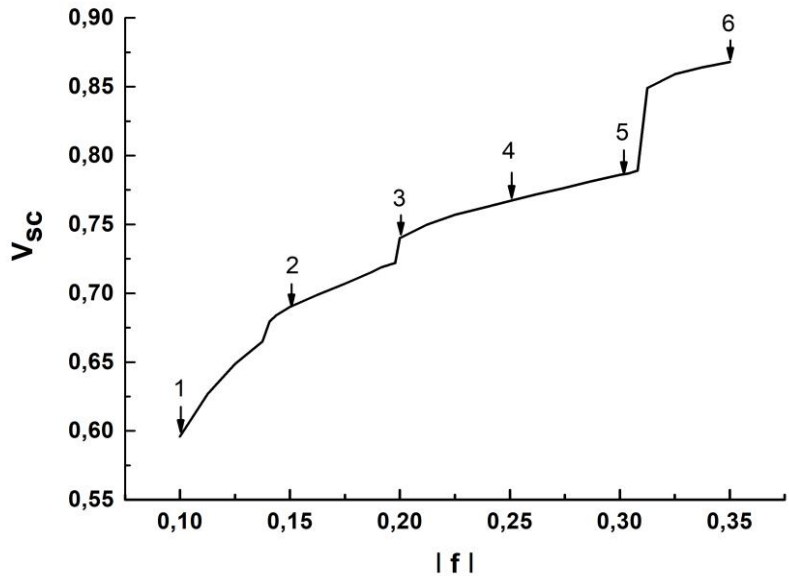

**Fig.8.** *A step-like dependence of the complex velocity on driving force strength. Arrows show velocity values corresponding to the complex profiles presented in Fig.7.*

## 7. Summary

Thus we have studied the nonstationary dynamics and interactions of topological solitons (kinks) in one-dimensional systems with a strong dispersion. An analytical approach has been proposed for investigation of dynamical features of a single kink motion, accompanied by emitting radiation and small-amplitude breathers. A collective coordinate ansatz has been also proposed for studying processes of soliton complex formation in relation to the strength of the dispersion, soliton velocity, and distance between solitons. The breather solution has been constructed in the small amplitude-limit and its internal oscillation and propagation in the dispersive medium have been investigated in detail. It has been shown that the theoretical results are in good relation with numerical simulations and explain them quantitatively. It is demonstrated that stable bound soliton states with complex internal structure can propagate in a dissipative medium owing to their stabilization by external forces.

The results obtained can be used for explanation and description of new effects in the dynamics of topological solitons in highly dispersive media – in particular, dislocations in nonideal lattices, fluxons in Josephson junction systems, and magnetic domain walls in anisotropic magnets.

### Acknowledgement

This work is dedicated to the memory of Arnold Markovich Kosevich, who was our teacher in science and life. We are grateful for a support to a joint scientific project No.24-02-a of NAS of Ukraine and RFBR, and to joint French-Ukrainian project in the framework of scientific cooperation between NAS of Ukraine and CNRS of France.